\documentclass[12pt]{article}
\usepackage[pdftex]{graphicx}               
\begin{document}
\section*{EFFECTS OF ROTATION ON PULSAR\\ RADIO PROFILES}
\vskip 0.5 cm
Dinesh Kumar and R. T. Gangadhara\\
Indian Institute of Astrophysics, Bangalore - 560034
\vskip 1cm
\noindent {\bf Abstract:} We have developed a model for pulsar
polarization by taking into account of viewing geometry, rotation and
modulation of the emission region. By solving for the plasma dynamics,
we deduced the expressions for curvature radiation electric field in
time as well as frequency domains. We show that both the
`antisymmetric' and `symmetric' types of circular polarization are
possible due to the combined effect of aberration, viewing geometry
and modulation.  We argue that aberration combined with modulation can
introduce `kinky' pattern into the PPA traverses.
\section{Introduction}
Pulsars are fast rotating, highly magnetized neutron stars which sweep
the radiation beam towards the observer once in a rotation
period. Pulsar radio emission is believed to be coherent curvature
radiation emitted by the relativistic plasma bunches streaming out
`force-freely' along the open field lines of the super-strong magnetic
field, the geometry of which is assumed to be predominantly dipolar.
In an inertial observer's frame (rotating case), the net velocity of
the plasma bunches will be offset from the tangents of their
associated field lines due to the effect of co-rotation, and hence the
emission beam gets aberrated towards the direction of rotation 
(Gangadhara, R.~T. 2005, ApJ, 628, 923).  At
any rotation phase $\phi'_{m}$, observer receives radiation from all
the emission points for which particle's velocity make an angle $\leq
1/\gamma$ with the observer's sight line, where $\gamma$ is the
Lorentz factor for plasma bunches.  The morphology of radio profiles
of pulsars are believed to be strongly influenced by the rotation
effects such as aberration and retardation (A/R).  A relativistic
pulsar emission model is being developed by taking into account of the
detailed geometry of emission region and aberration.

\section{Radiation electric field}
By computing the velocity and acceleration of the relativistic plasma,
estimated the electric field of radiation, and obtained the spectral
distribution by taking Fourier transform. The Stokes parameters: $I,$
$Q,$ $U,$ and $V$ are being used to specify the polarization state of
the radiation. The modeling of Stokes parameters for pulsar radio
emission is very important as they have been found to best in
exploring the association between the polarization of the observed
radiation, the emission mechanism, and the geometry of the emission
region (Gangadhara, R.~T. 2010, ApJ, 710, 29).  The parameter $I$
defines the total intensity, $Q$ and $U$ jointly define the linear
polarization $L = \sqrt{Q^{2} + U^{2}}$ and position angle of the
radiation field $\psi = ({1/ 2}) \tan^{-1}(U/Q)$, and $V$ defines the
circular polarization.
\begin{figure}[htp]
\centering
\includegraphics[width=12cm,bb=0 0 375 430]{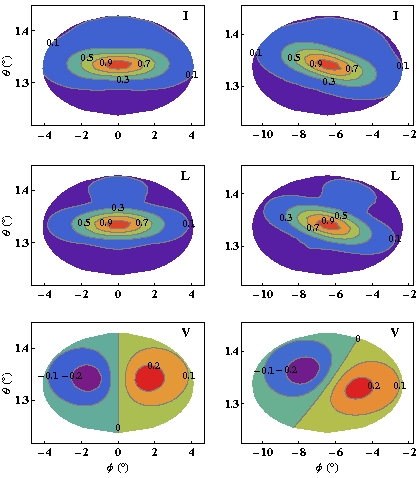}
\caption{The simulation results showing the spectral emission 
  from the beaming region in the non-rotating (left column panels) and the
  rotating (right column panels) cases of pulsar.}
\end{figure}

\begin{figure}[htp]
\centering
\includegraphics[width=12cm,bb=0 0 375 430]{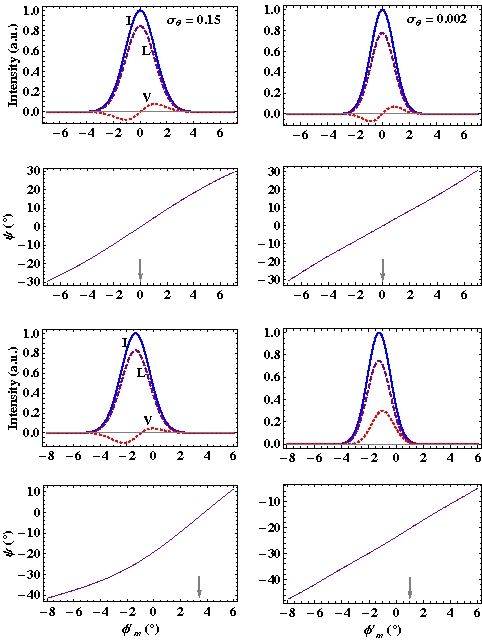}
\caption{Simulated pulse profiles: The panels in the top two rows are
  for non-rotating case, and bottom two for the rotating. Chosen
  $f_{0}=1$, $\theta_{P}=1.2^{\circ}$, $\phi_{P}=0^{\circ}$, and
  $\sigma_{\phi}=0.15$. In each panel $I$ (solid blue curves), $L$
  (dashed purple curves), and $V$ (dotted red curves) are normalized
  with the respective peak intensity. The continuous pink curve
  present the polarization angle ($\psi$) swing. Arrows mark the phase
  of intensity peak and the inflection point of $\psi$.}
\end{figure}

\section{Simulation of polarization profiles}
Using the parameters magnetic axis inclination angle
$\alpha=10^{\circ},$ sight line impact parameter $\sigma=2^{\circ},$
$\phi'_{m}=0^{\circ},$ emission altitude $r_{n}=r/r_{LC}=0.02,$ pulsar
period $P=1$~s, $\gamma=400$, and observational frequency $\nu=600$
MHz, simulated to obtain the parameters: $I,$ $L,$ and $V,$ within the
beaming region. Their contour plots are given in magnetic coordinates
($\theta,~\phi$)-plane in Figure~1. Emission altitude is $r,$ and
light cylinder radius is $r_{LC}=c~P/2\pi,$ where $c$ is the speed of
light.  For comparison presented both the cases: non-rotating (left
column panels), and rotating (right column panels). The parameters
$I,$ $L,$ and $V$ in both non-rotating and rotating cases are
normalized with the corresponding maximum value of $I.$ The contour
levels are marked on the respective contours. Due to aberration,
contour patterns in the rotating case gets rotated in the
$(\theta,~\phi)$ plane as compared to the corresponding ones in the
non rotating case. Hence an asymmetry in the strength of emission
occurs within the beaming region bounded by the ranges of $\theta$ and
$\phi$ allowed by the geometry. This rotation of pattern is
responsible for the wide diversity of circular polarization and
position angle swing.

 The shapes of
pulsar intensity profile indicate that the entire polar cap do not
radiate uniformly. In general, pulsar average radio profiles consist
of many components: central core emission surrounded by concentric
conal emissions. Hence the emission region can be modeled by assuming
Gaussian sources distributed within the emission region.  Hence a
modulation function for a pulse component may defined as
$$
f(\theta,\phi) = f_{0} \exp[ -(\theta-\theta_{P})^{2}/\sigma_{\theta}^{2} 
                -(\phi-\phi_{P })^{2}/\sigma_{\phi}^{2}],
$$
where $\theta_{P}$ and $\phi_{P}$ are the peak locations and $f_{0}$
is the amplitude. The parameters $\sigma_{\theta} =
w_{\theta}/(2\sqrt{ln 2})$ and $\sigma_{\phi} = w_{\phi}/(2\sqrt{ln
  2})$, where $w_{\theta}$ and $w_{\phi}$ are the full width at
half-maxima (FWHM).

By considering a Gaussian modulation having peak at
($\theta_{P},~\phi_{P})=(1.2^{\circ},~0^{\circ})$ and using the same
parameters used for Figure 1, simulated the polarization profiles and
presented in Figure 2. To see the combined effect of aberration and
modulation, considered two cases of modulation: $\sigma_{\theta}=0.15$
and 0.002, and kept $\sigma_{\phi}=0.15$ constant. From top the panels
in first and second row are for non-rotating case, and third and last
row panels for the rotating case.  In all the three cases of
$\sigma_{\theta}$, the intensity profiles in the rotating case are
found to be shifted to the earlier phase whereas the polarization
position angle profiles shifted to the later phase due to aberration.
The phase shifts of $I$ peak in the cases of $\sigma_{\theta}=0.15$
and 0.002 are found to be $1.36^{\circ}$ and $1.24^{\circ}$,
respectively. The phase shift of $I$ peak is found to decrease
with decreasing $\sigma_{\theta}$ due to increase in the steepness of
the modulation. In the non-rotating case, the circular $V$ is always
antisymmetric with sign reversal from negative to positive
irrespective of the modulation parameters, and in any case negative
and positive circulars are of equal magnitude. But in the rotating
case, due to combined effect of aberration and modulation, $V$ is
antisymmetric type in the case of $\sigma_{\theta}=0.15$ but
asymmetric in the positive and negative portions. In the extreme case
of $\sigma_{\theta}=0.002$, $V$ becomes symmetric type, i.e., only
strong positive circular survived. The simulations for the first time
reproduced this important result on the generation of symmetric type
circular polarization.

In all the the cases, the linear polarization $L$ almost follows the
total intensity except for its lower value due to incoherent addition
of Stokes parameters within the beaming region. The polarization
position angle swing is counter clockwise, and the position angle
inflection points (indicated by arrows) are found to be shifted to
later phases by $3.42^{\circ}$ and $1.00^{\circ}$ in the rotating
case.  The phase shifts of the position angle inflection point is
found to decrease with decreasing $\sigma_{\theta}$ due to combined
effect of aberration and modulation.
\section{Conclusion}
Developed a relativistic model of pulsar radio emission and
polarization.  This is for the first time we are able to explain the
complete polarization state of the coherent curvature radiation by
taking into account of the effect of rotation. Our model predicts both
the symmetric as well as antisymmetric type of circular polarization
is possible within the frame work of coherent curvature radiation.
\vskip 1 cm
\noindent
{\bf Acknowledgment:}
We thank J. L. Han and Pengfei Wang for stimulating discussions.
\end{document}